\documentclass[letterpaper, 10pt, conference]{IEEEconf}      % Use this line for a4
\IEEEoverridecommandlockouts                             
\usepackage{graphics} % for pdf, bitmapped graphics files
\usepackage{epsfig} % for postscript graphics files
\usepackage{mathptmx} % assumes new font selection scheme installed
\usepackage{times} % assumes new font selection scheme installed
\usepackage{amsmath} % assumes amsmath package installed
\usepackage{amssymb}  % assumes amsmath package installed
\usepackage{eucal}
\usepackage{xcolor}

\usepackage{amsthm}
\newtheorem{thm}{Theorem}

\newtheorem{definition}{Definition}

\title{\LARGE \bf
A Coherent LQG approach to Quantum Equalization}

\author{Rebbecca TY Thien, Shanon L. Vuglar and Ian R. Petersen% <-this % stops a space
\thanks{This work was supported by the Australian Research Council under grant DP210101938. It was also supported by the Office of Naval Research Global under agreement number N62909-19-2129.}% <-this % stops a space
\thanks{Rebbecca TY Thien is with the School of Engineering, The Australian National University, Canberra ACT 2601, Australia
        {\tt\small rebbecca.thien@anu.edu.au}}%
\thanks{Shanon L. Vuglar is with the College of STEM and Health Professions, John Brown University, Siloam Springs AR 72761, United States of America
        {\tt\small shanonvuglar@gmail.com}}%
\thanks{Ian R. Petersen is with the School of Engineering, The Australian National University, Canberra ACT 2601, Australia 
        {\tt\small i.r.petersen@gmail.com}}%
}
\begin{document}
\maketitle
\thispagestyle{empty}
\pagestyle{empty}
%%%%%%%%%%%%%%%%%%%%%%%%%%%%%%%%%%%%%%%%%%%%%%%%%%%%%%%%%%%%%%%%%%%%%%%%%%%%%%%%%%%%
\begin{abstract}
We propose a method to design a suboptimal, coherent quantum LQG controller to solve a quantum equalization problem. Our method involves reformulating the problem as a control problem and then designing a classical LQG controller and implementing it as a quantum system. Illustrative examples are included which demonstrate the algorithm for both active and passive systems, i.e., systems where the dynamics are described in terms of both position and momentum operators and systems with dynamics in terms of annihilation operators only.
\end{abstract}
%%%%%%%%%%%%%%%%%%%%%%%%%%%%%%%%%%%%%%%%%%%%%%%%%%%%%%%%%%%%%%%%%%%%%%%%%%%%%%%%
\section{INTRODUCTION}
Communication systems are necessary for transmitting information over long distances, however, this often results in degradation of the quality. The goal of equalization is to estimate the transmitted signal from the received signal, compensating for the effects of noise and distortion. This is typically done by designing a filter that maps the received signal to an estimate of the original signal~\cite{Gregory}. 

\begin{figure}[htp!] 
    \centering
    \includegraphics[width=6cm]{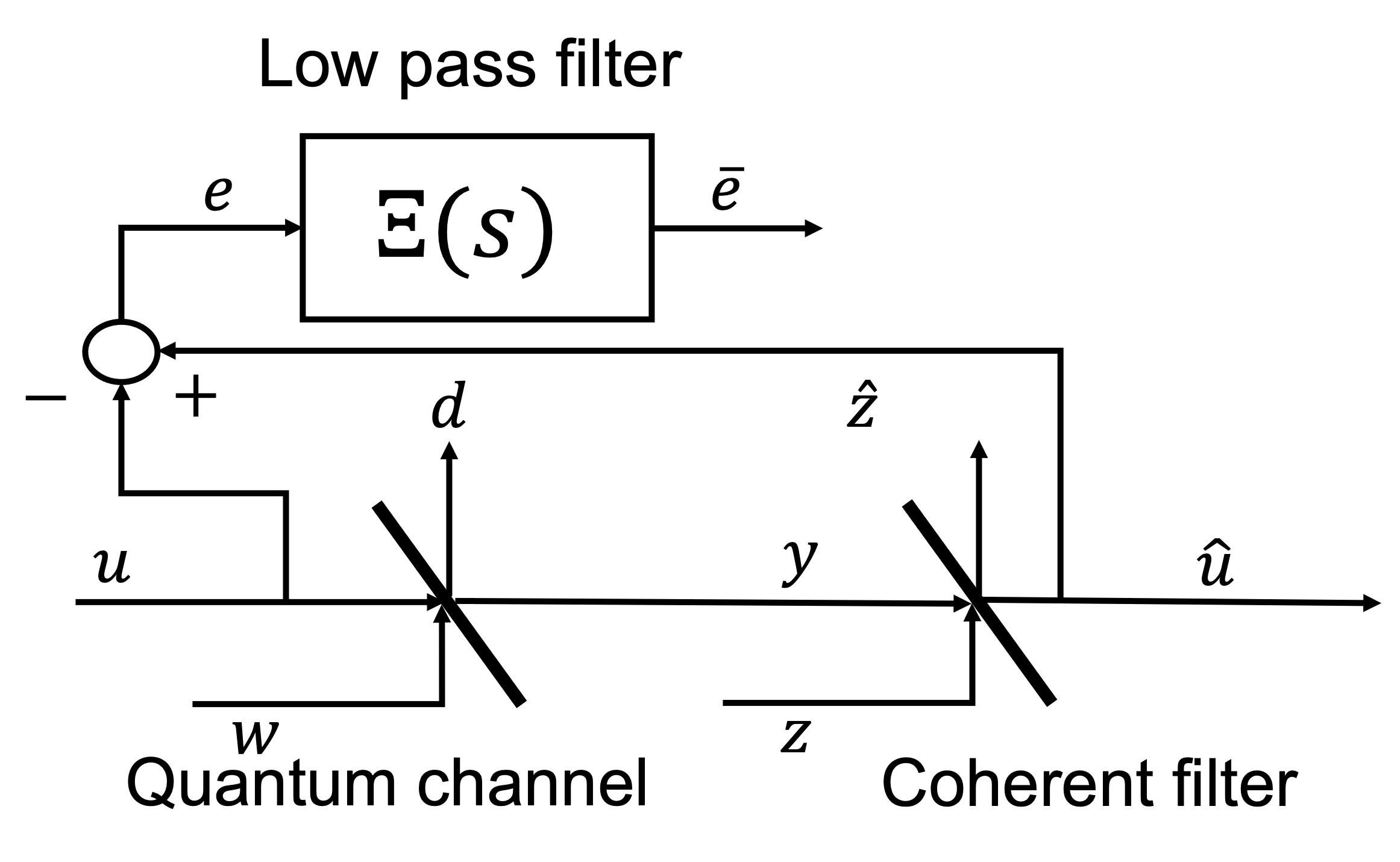}
    \caption{A quantum optical communication system consisting of two beam splitters acting as a channel and a filter, respectively.}
    \label{figintro}
\end{figure}

In the case of quantum communication systems, the laws of quantum mechanics limits their capacity to transfer information. Hence, the problem of correcting distortions in quantum communication systems is complex compared to its classical counterpart~\cite{UgrinovskiiJames1}. This problem is called the quantum equalization problem~\cite{UgrinovskiiJames}, and is depicted in Figure~(\ref{figintro}). 

Quantum linear systems are a class of quantum systems whose dynamics take the specific form of a set of linear quantum stochastic differential equations~(QSDEs). Such systems are common to the area of quantum optics~\cite{GardinerZoller, BachorRalph}, and \cite{WallsMilburn}. In general, a set of linear QSDEs need not correspond to a physically meaningful quantum system. To represent a physical quantum system, they must satisfy additional constraints; this leads to the notion of a physically realizable quantum system. This is discussed in~\cite{JamesNurdinPetersen}, \cite{VuglarPetersen}, and \cite{MaaloufPetersen}, where the authors derive necessary and sufficient conditions for such systems. 

The goal of feedback control of quantum systems is to achieve closed-loop properties, such as stability, robustness, and entanglement. Coherent quantum control is a type of feedback control in which the controller itself is also a quantum system. This type of control has attracted considerable interest in recent years, since the use of a quantum controller may lead to an improved performance of the system, ease of implementations, or both~\cite{NurdinJamesPetersen, XiangPetersenDong}, and \cite{DongPetersen}. 

In this work, we propose a novel approach to solving the equalization problem by converting it into a coherent Linear Quadratic Gaussian~(LQG) problem. This approach has several advantages over existing methods~\cite{UgrinovskiiJames}, and it provides a simple and systematic way to solve the equalization problem for both passive and active systems. Using our proposed approach, we can design a controller~(fulfilling the role of an equalization filter) that optimizes the performance of the communication system while minimizing the impact of noise and distortion. The main difference between our work and~\cite{UgrinovskiiJames}, is that we are using the coherent LQG control, while~\cite{UgrinovskiiJames} uses a $H_{\infty}$-like methodology. It is not possible to do a direct comparison between $H_{\infty}$ control methods and LQG control methods, since their performance indices are measuring different quantities~\cite{Petersen1}. 

The main contribution of this work is threefold. Firstly, we propose an algorithm that solves the equalization problem for a passive system. Secondly, we extend our approach to solve the equalization problem for an active system, which is an extension of existing methods that only work for passive systems. Lastly, we demonstrate the practical relevance of our proposed method by giving an application in a real-world scenario.

Our proposed approach adapts results from~\cite{VuglarPetersen}, \cite{MaaloufPetersen}, and \cite{ThienVuglarPetersen}. By converting the equalization problem into a coherent LQG problem, we can then design filters that optimize the performance of communication systems while minimizing the impact of noise and distortion. The remainder of the paper proceeds as follows: in Section~\ref{linearquantumsystem} and~\ref{PR}, we describe the quantum linear system models under consideration and define the corresponding notion of physical realizability, respectively. The latter section also includes some relevant previous results. Then, we formulate our problem in Section~\ref{problem formulation} and propose our algorithm in Section~\ref{algo}. Examples are given in Section~\ref{example} followed by a conclusion and future work in Section~\ref{conclusion}.
%%%%%%%%%%%%%%%%%%%%%%%%%%%%%%%%%%%%%%%%%%%%%%%%%%%%%%%%%%%%%%%%%%%%%%%%%%%%%%%%
\section{Linear Quantum Systems}\label{linearquantumsystem}
We consider both passive and active linear quantum systems. Here, passive means that the system is defined in terms of annihilation operators only, while active means that the system is defined in terms of annihilation and creation operators. For the active system, we use position and momentum operators for convenience, so that we can directly use the results of~\cite{ThienVuglarPetersen}.

\subsection{Passive Quantum Systems}
Passive quantum systems are a class of systems that can be described using non-commutative or quantum probability theory~\cite{BoutenHandelJames}. In particular, the systems under consideration are described in terms of complex annihilation operators satisfying the linear quantum stochastic differential equations~(QSDEs)
    \begin{equation} \label{passivemodel}
        \begin{split}
            da(t) & = Fa(t)dt+Gdw(t); \\
            dy(t) & = Ha(t)dt+Jdw(t) \\
        \end{split}
    \end{equation}
where $F \in \mathbb{C}^{n \times n}$, $G \in \mathbb{C}^{n \times n_w}$, $H \in \mathbb{C}^{n_y \times n}$, $J \in \mathbb{C}^{n_y \times n_w}$~($n$, $n_y$, $n_w$ are positive integers). Here $a(t)$ $=$$[a_1(t) \dots a_n(t)]^T$ is a vector of annihilation operators on an underlying Hilbert space~\cite{JamesNurdinPetersen, Petersen1}.

The quantity $w$ describes the input variables and is assumed to admit the decomposition 
    \begin{equation*}
        dw(t)=\beta_w(t)dt+d\Tilde{w}(t)
    \end{equation*}
where $\Tilde{w}$ is the noise part of $w(t)$ and $\beta_w(t)$ is an adapted process~\cite{HudsonParthasarathy, Parthasarathy,Belavkin}.The noise $\Tilde{w}(t)$ is an operator-valued process with a vector of quantum Weiner processes with a quantum Ito table
    \begin{equation*}
        d\Tilde{w}(t)d\Tilde{w}^\dagger (t)=F_{\Tilde{w}}dt
    \end{equation*}
where $F_{\Tilde{w}}$ is a nonnegative Hermitian matrix \cite{HudsonParthasarathy, Parthasarathy}, and \cite{Belavkin}. Here, the notation $\dagger$ represents the adjoint transpose of a vector of operators. It is also assumed that the following communitation relations hold ofr the noise components:
    \begin{equation*}
        [d\Tilde{w}(t), d\Tilde{w}(t)^\dagger]\triangleq
        d\Tilde{w}(t)d\Tilde{w}(t)^\dagger - (d\Tilde{w}(t)d\Tilde{w}(t)^T)^T =T_wdt
    \end{equation*}
where $T_w$ is a Hermitian commutation matrix.
\subsection{Active Quantum Systems}
An active quantum system is a system where the dynamics are described in terms of annihilation and creation or position and momentum operators. It can be described by the following linear quantum stochastic differential equations~(QSDEs)~\cite{JamesNurdinPetersen,VuglarPetersen, HudsonParthasarathy, Parthasarathy}, and \cite{Belavkin}:
    \begin{equation} \label{classicmodel}
        \begin{split}
            dx(t) & = Ax(t)dt+Bdw(t); \\
            dy(t) & = Cx(t)dt+Ddw(t) \\
        \end{split}
    \end{equation}
where $A, B, C$ and $D$ are real matrices in $\mathbb{R}^{n \times n}$, $\mathbb{R}^{n \times n_w}$, $\mathbb{R}^{n_y \times n}$ and $\mathbb{R}^{n_y \times n_w}$~($n, n_w, n_y$ are even positive integers), respectively. Moreover, $x(t)=[x_1(t) ... x_n(t)]$ is a column vector of self-adjoint, possibly non-commutative, system variables. 
 
Equations~(\ref{classicmodel}) must also preserve certain \textit{commutation relations} as follows:
    \begin{equation} \label{ccr}
        [x_j(t), x_k(t)]=x_j(t)x_k(t)-x_k(t)x_j(t)=2i \Theta_{jk}
    \end{equation}
where $\Theta$ is a real skew-symmetric matrix with components $\Theta_{jk}$ where $j,k = 1, ..., n$ and $i=\sqrt{-1}$ in order to represent the dynamics of a physically meaningful quantum system.

The \textit{commutation relations} (\ref{ccr}) are said to be \textit{canonical} if 
    \begin{equation} \label{theta}
        \Theta_m = \mathrm{diag}(J_\Theta, J_\Theta, ..., J_\Theta)
    \end{equation}
    where $J_\Theta$ denotes the real skew-symmetric $2 \times 2$ matrix
    \begin{equation*}
        J_\Theta =\begin{bmatrix} 
        0 & 1 \\
        -1 & 0
        \end{bmatrix}
    \end{equation*} 
and the ``$\mathrm{diag}$'' notation indicates a block diagonal matrix assembled from the given entries. Here $m$ denotes the dimension of the matrix $\Theta_m$.

The vector quantity $w$ describes the input signals and is assumed to admit the decomposition 
    \begin{equation*}
        dw(t)=\beta_w(t)dt+d\Tilde{w}(t)
    \end{equation*}
where the self-adjoint, adapted process $\beta_w(t)$ is the signal part of $dw(t)$ and $d\Tilde{w}$ is the noise part of $dw(t)$ \cite{HudsonParthasarathy, Parthasarathy,Belavkin}.The noise $\Tilde{w}(t)$ is a vector of self-adjoint quantum noises with Ito table
    \begin{equation*}
        d\Tilde{w}(t)d\Tilde{w}^T(t)=F_{\Tilde{w}}dt
    \end{equation*}
where $F_{\Tilde{w}}=S_{\Tilde{w}}+T_{\Tilde{w}}$ is a nonnegative Hermitian matrix \cite{Belavkin,Parthasarathy} with $S_{\Tilde{w}}$ and $T_{\Tilde{w}}$ are real and imaginary, respectively. 
In this paper, we will assume $F_{\Tilde{w}}$ is of the form $F_{\Tilde{w}} = I + i\Theta$ where $\Theta$ is of the form~(\ref{theta}).

In this work, we consider a special case of~(\ref{classicmodel}):
    \begin{equation} \label{specialcase}
        \begin{split}
            dx(t) & = Ax(t)dt+B_udu(t)+B_{v}dv(t); \\
            dy(t) & = Cx(t)dt+dv(t); \\
        \end{split}
    \end{equation}
see also \cite{JamesNurdinPetersen, VuglarPetersen, ThienVuglarPetersen}.
Here, $dw(t)$ from~(\ref{classicmodel}) has been partitioned into the signal input $du(t)$~(a column vector with $n_u$ components) and the direct feed through quantum vacuum noise input $dv(t)$. 
We can regard such a quantum system as a coherent controller in a coherent quantum feedback control system; e.g., see~\cite{JamesNurdinPetersen, VuglarPetersen}.
%%%%%%%%%%%%%%%%%%%%%%%%%%%%%%%%%%%%%%%%%%%%%%%%%%%%%%%%%%%%%%%%%%%%%%%%%%%%%%%%

\section{Physical Realizability}\label{PR}
\subsection{Passive Quantum Systems}
In~\cite{MaaloufPetersen1}, the notion of physical realizability is developed based around the concept of a complex open quantum harmonic oscillator. We consider a passive quantum plant described by the following equations which are in terms of annihilation operators:
\begin{equation}\label{passivedefplant}
    \begin{split}
        da(t) & = Fa(t)dt + 
        \begin{bmatrix}
            G_0 & G_1 & G_2
        \end{bmatrix} \\
        & \begin{bmatrix}
            dv(t)^T & dw(t)^T & du(t)^T
        \end{bmatrix}^T; \\
        dz(t) & = H_1a(t)dt +J_{12}du(t); \\
        dy(t) & = H_2a(t)dt + 
        \begin{bmatrix}
            J_{20} & J_{21} & 0_{n_y \times n_y}
        \end{bmatrix} \\
        & \begin{bmatrix}
            dv(t)^T & dw(t)^T & du(t)^T
        \end{bmatrix}^T
    \end{split}
\end{equation}
where $\mathbb{F} \in \mathbb{C}^{n \times n}$, $\mathbb{G}_0 \in \mathbb{C}^{n \times n_v}$, $\mathbb{G}_1 \in \mathbb{C}^{n \times n_w}$, $\mathbb{G}_2 \in \mathbb{C}^{n \times n_u}$, $\mathbb{H}_1 \in \mathbb{C}^{n_z \times n_w}$, $\mathbb{J}_{12} \in \mathbb{C}^{n_z \times n_u}$,  $\mathbb{H}_2 \in \mathbb{C}^{n_y \times n_n}$, $\mathbb{J}_{20} \in \mathbb{C}^{n_y \times n_v}$ and $\mathbb{J}_{21} \in \mathbb{C}^{n_y \times n_w}$.

Similarly a controller is defined as follows:
\begin{equation}\label{passivedefcontroller}
    \begin{split}
        d\xi (t) & = F_c\xi (t) dt + \begin{bmatrix}
            G_{c_0} & G_{c_1} & G_c
        \end{bmatrix} 
        \begin{bmatrix}
            dw_{c_0}(t) \\ dw_{c_1}(t) \\ dy(t)
        \end{bmatrix}\\
        du(t) & = H_c\xi (t)dt+dw_{c_0}
    \end{split}
\end{equation}
where $\xi (t) = [\xi_1(t) \ldots \xi_n(t) ]^T$ is a vector of controller annihilation operator variables. We now define the notion of physically realizable for this class of systems.
% defined by $n_w$ measurement channels coupled via the operator $L$ $=$ $\Lambda a$~($\Lambda$ is a complex $n_w$ $\times$ $n$ matrix) and a Hamiltonian $\mathfrak{H}$ $=$ $a^\dagger Ma$~(where $M$ is a $n \times n$ complex Hermitian matrix).
% \begin{definition}~\cite[Definition 2.3]{MaaloufPetersen1}
% An annihilation-operator linear quantum system of the form~(\ref{passivemodel}) is said to be physically realizable if it satisfies the generalized commutation relations
%     \begin{equation*}
%         F\Theta + \Theta F^\dagger +GG^\dagger = 0
%     \end{equation*}
% with $\Theta$ $=$ $\Theta^\dagger$ $\>$ $0$, $T_w$ $=$ $I$ and moreover
%     \begin{equation*}
%         G = -\Theta \Lambda^\dagger; \quad H = \Lambda; \quad J=I.
%     \end{equation*}
% \end{definition}
% \begin{thm}~\cite[Theorem 2.2]{MaaloufPetersen1}
% An annihilation-operator linear quantum system of the form~(\ref{passivemodel}) is physically realizable if and only if there exists a matrix $\Theta$ $=$ $\Theta^\dagger$ $>$ $0$ such that
% \begin{equation*}
%         F\Theta + \Theta F^\dagger +\Theta H^\dagger H\Theta^\dagger = 0; \quad J = I.
%     \end{equation*}
% \end{thm}

\begin{definition}\cite[Definition 3.1]{MaaloufPetersen}
    The matrices $F_c$, $G_c$, $H_c$
%     \begin{equation*} 
%         \begin{split}
%             F_c & = F+G_2H_c-G_cH_2+(G_1-G_cJ_{21})G_1^\dagger X; \\
%             G_c & = (I-YX)^{-1}(YH_2^\dagger + G_1J_{21}^\dagger)E_2^{-1}; \\
%             H_c & = -E_1^{-1}(g^2G_2^\dagger X+H_{12}^\dagger H_1)
%         \end{split}
%     \end{equation*}
% where $X$ and $Y$ is a stabilizing solution for the following complex algebraic Riccati equations~(ARE): 
%  \begin{equation*} 
%         \begin{split}
%             & (F-G_2E_1^{-1}J_{12}^\dagger H_1)^\dagger X + X(F-G_2E_1^{-1}J_{12}^\dagger H_1) \\
%             & +X(G_1G_1^\dagger - g^2G_2E_1^{-1}G_2^\dagger)X \\
%             & +g^{-2}H_1^\dagger(I-J_{12}E_1^{-1}J_{12}^\dagger)H_1 = 0; \\
%             & (F-G_1J_{21}^\dagger E_2^{-1} H_2)Y + Y(F-G_1J_{21}^\dagger E_2^{-1}H_2)^\dagger \\
%             & +Y(g^{-2}H_1^\dagger H_1-H_2^\dagger E_2^{-1} H_2)Y \\
%             & G_1(I-J_{21}^\dagger E_2^{-1}J_21)G_1^\dagger= 0
%         \end{split}
%     \end{equation*}
are said to define a physically realizable controller of the form~(\ref{passivedefcontroller}) if there exists matrices $G_{c_0}$, $G_{c_1}$, $H_{c_1}$ and $H_{c_2}$ such that the quantum system of the form~(\ref{passivemodel})
\begin{equation} \label{passivedefcl}
    \begin{split}
        d\xi (t) & = F_c\xi (t) dt + \begin{bmatrix}
            G_{c_0} & G_{c_1} & G_c
        \end{bmatrix}\begin{bmatrix}
            dw_{c_0} \\
            dw_{c_1}\\
            dy
        \end{bmatrix}; \\
        \begin{bmatrix}
            du \\ du_1 \\ du_2 
        \end{bmatrix} & \begin{bmatrix}
            H_c \\ H_{c_1} \\ H_{c_2} 
        \end{bmatrix} \xi (t)dt + 
        \begin{bmatrix}
            I & 0 & 0 \\
            0 & I & 0 \\
            0 & 0 & I
        \end{bmatrix}
        \begin{bmatrix}
            dw_{c_0} \\
            dw_{c_1}\\
            dy
        \end{bmatrix}
    \end{split}
\end{equation}
is physically realizable when $T_y$ $=$ $J_{20}T_vJ_{20}^\dagger$ $+$ $J_{21}T_wJ_{21}^\dagger$ $=$ $I$.
\end{definition}
\begin{thm}\cite[Theorem 3.2]{MaaloufPetersen} \label{Maaloufmethod}
    Suppose the matrices $F_c$, $G_c$, and $H_c$ are such the corresponding system is minimal~\cite{MaaloufPetersen1}. Then the matrices $F_c$, $G_c$, and $H_c$ define a physically realizable controller of the form~(\ref{passivedefcontroller}) if and only if $F_c$ is Hurwitz and 
    \begin{equation*}
        \|H_c(sI-F_c)^{-1}G_c\|_\infty \leq 1
    \end{equation*}
i.e., the corresponding system is bounded real~\cite{MaaloufPetersen1}. In this case, the matrices $G_{c_1}$ and $H_{c_1}$ in~(\ref{passivedefcl}) can be taken as zero.
\end{thm}

\subsection{Active Quantum Systems}
In~\cite{JamesNurdinPetersen}, the notion of physical realizability is based on the concept of an open quantum harmonic oscillator. The following formally defines physical realizability for the more general case of active quantum systems.
    \begin{definition}~\cite[Definition 3.1]{JamesNurdinPetersen}
    The system~(\ref{classicmodel}) is said to be physically realizable if $\Theta$ is canonical and there exists a quadratic Hamiltonian operator \begin{math}\mathcal{H}=(1/2)x(0)^TRx(0) \end{math}, where $R$ is a real symmetric \begin{math}n \times n \end{math} matrix, and a coupling operator \begin{math} \mathcal{L} = \Lambda x(0) \end{math}, where $\Lambda$ is a complex-valued $\frac{n_w}{2} \times n$ coupling matrix such that matrices $A, B, C,$ and $D$ are given by 
        \begin{subequations} \label{oqhs}
            \begin{align}
                A & = 2 \Theta (R + \Im{(\Lambda^\dagger \Lambda)}) \\
                B & = 2i\Theta[-\Lambda^\dagger \quad \Lambda^T]\Gamma \\ 
                C & = P^T 
                    \begin{bmatrix} 
                        \Sigma_{n_y} & 0 \\
                        0 & \Sigma_{n_y} 
                    \end{bmatrix}
                    \begin{bmatrix} 
                    \Lambda+\Lambda^\# \\
                    -i\Lambda+i\Lambda^\# \end{bmatrix} \\
                    D & = [I_{n_y \times n_y} \quad 0_{n_y \times (n_w-n_y)}].
            \end{align}
        \end{subequations}
    Here 
        \begin{equation*}
            \begin{split}
                \Gamma & =P_{N_w}diag_{N_w}(M); \\
                M & =\frac{1}{2}    
                    \begin{bmatrix} 
                        1 & i \\
                        1 & -i 
                    \end{bmatrix}; \\
                \Sigma_{N_y} & =[I_{N_y \times N_y} \quad 0_{N_y \times(N_w-N_y)}]; \\
                P_{N_w}(a_1, a_2,..., a_{2N_w})^T & =(a_1,..., a_{2N_w-1}, a_2,..., a_{2N_w})^T;
            \end{split}
        \end{equation*} 
    and diag$(M)$ is an appropriately dimensioned square block diagonal matrix with each diagonal block equal to the matrix $M$. 
    Note that the permutation matrix $P$ has the unitary property $PP^T=P^TP=I$ and $N_w=n_w/2$ and $N_y=n_y/2$.
    \end{definition}
    
The following theorem~\cite{JamesNurdinPetersen} gives necessary and sufficient conditions for the physical realizability of our system~(\ref{specialcase}).
    \begin{thm} \label{theoremPR}~\cite[Theorem 3.4]{JamesNurdinPetersen}
        The system~(\ref{specialcase}) is physically realizable if and only if
            \begin{equation*} 
                \begin{split}
                    A\Theta_n + \Theta_n A^T + B_{v}\Theta_{n_v} B_{v}^T + B_u\Theta_{n_u} B_u^T & = 0; \\
                    B_{v} \begin{bmatrix} 
                    I_{n_y \times n_y} \\
                    0_{(n_w-n_y)\times n_y}
                    \end{bmatrix} & = \Theta C^T diag(J); 
                \end{split}
            \end{equation*}
        where $\Theta_n, \Theta_{n_v}$ and $\Theta_{n_u}$ are all defined as in~(\ref{theta}) but may be of different dimensions.
\end{thm}
Here $(.)^{\dagger}$ denotes the complex conjugate transpose of a matrix while $(.)^{\#}$ denotes the complex conjugate of a matrix.

In our work, we consider the Linear Matrix Inequality~(LMI) version of the physical realizability which is similar to the approach in~\cite{JamesNurdinPetersen, VuglarPetersen} but reformulated into an LMI problem~\cite[Section 4]{ThienVuglarPetersen}.
%%%%%%%%%%%%%%%%%%%%%%%%%%%%%%%%%%%%%%%%%%%%%%%%%%%%%%%%%%%%%%%%%%%%%%%%%%%%%%%%%%%%%%%%%%%%%%%%%
\section{Problem Formulation} \label{problem formulation}
The problem formulation described in this work is similar to~\cite{NurdinJamesPetersen, VuglarPetersen}, with some minor differences. Suppose we have a quantum plant described by the following QSDEs which are a special case of~(\ref{specialcase}): 
    \begin{equation} \label{plant}
        \begin{split}
            dx(t) & = Ax(t)dt+B_{\hat{u}}d\hat{u}(t)+B_{w_1}dw_1(t); \\
            dy(t) & = Cx(t)dt+D_{w_1}dw_1(t)
        \end{split}
    \end{equation}
where the vector $dw_1$$=$$[du \quad dw]^T$.
This quantum plant~(\ref{plant}) can be obtained through the combination of the dynamics of the quantum channel:
    \begin{equation*} \label{qchannel}
        \begin{split}
            dx(t) & = Ax(t)dt+B_udu(t)+B_wdw(t); \\
            dy(t) & = Cx(t)dt+D_udu(t)+D_wdw(t)
        \end{split}
    \end{equation*}
and the low-pass filter:
    \begin{equation} \label{lowpassfilter}
        \begin{split}
            dx_f(t) & = A_fx_f(t)dt+B_{\hat{u}}d\hat{u}(t)+B_udu(t); \\
            \Bar{e}(t) & = x_f(t)
        \end{split}
    \end{equation}
as shown in Figure~\ref{figintro}. 

Also, suppose that we wish to minimize an infinite horizon quadratic cost function: 
\begin{equation} \label{cost}
    J_{cost}=\lim_{t_f\to\infty}\frac{1}{t_f} \int_{0}^{t_f}\big \langle \bar{e}(t)^TR_1\bar{e}(t)+\mu \hat{u}(t)^T R_2 \hat{u}(t)) \big \rangle dt.
\end{equation}
The low pass filter~(\ref{lowpassfilter}) is introduced so that the cost function~(\ref{cost}) will be well defined. This is justified since in practice, the equalization filter only needs to work over a finite bandwidth rather than an infinite bandwidth. 
%where $\bar{e}$ is obtained by passing the signal $e$~($e$ $=$ $d\hat{u} - du$) through a low-pass filter. 

The problem is as follows: given a quantum plant of the form~(\ref{plant}), design a classical LQG controller of the form
    \begin{equation} \label{classiccontroller}
        \begin{split}
            dx(t) & = A_k x_k(t)dt+B_y dy(t); \\
            d\hat{u}(t) & = C_k x_k(t)dt.
        \end{split}
    \end{equation}
that minimizes the cost function~(\ref{cost}), which then be implemented as a physically realizable LQG quantum controller of the form
    \begin{equation} \label{controller}
        \begin{split}
            dx(t) & = A_kx_k(t)dt+B_ydy(t)+B_{v_1}d_{v_1}(t)+B_{v_2}d_{v_2}(t); \\
            d\hat{u}(t) & = C_kx_k(t)dt+dv_1(t).
        \end{split}
    \end{equation}
%%%%%%%%%%%%%%%%%%%%%%%%%%%%%%%%%%%%%%%%%%%%%%%%%%%%%%%%%%%%%%%%
\section{Algorithm}\label{algo}
The main idea of our algorithm is to design a classical LQG controller and then use the results in~\cite{MaaloufPetersen}~(or \cite{ThienVuglarPetersen}) to implement this controller as a physically realizable quantum system.

To begin with, we form a classical LQG problem. Consider the quantum plant~(\ref{plant}) and the classical controller~(\ref{classiccontroller}).
This classical LQG problem can be solved in the usual manner~\cite[Theorem 5]{KwakSivan}. The solution is the controller~(\ref{classiccontroller}) with 
    \begin{equation*} 
        \begin{split}
            A_k & = A-KC-B_{\hat{u}}+KD_{\hat{u}}F; \\
            B_y & = K; \\
            C_k & = -F.
        \end{split}
    \end{equation*}
The matrices $F$ and $K$ can be obtained as follows:
    \begin{equation*} 
        F = R_2^{-1}B_{\hat{u}}^TP
    \end{equation*}
where $P \geq 0$ is the solution to the ARE:
    \begin{equation*} 
        A^TP+PA+PB_{\hat{u}}R_2^{-1}B_{\hat{u}}^TP+R_1=0,
    \end{equation*}
and 
\begin{equation*} 
        K = (QC^T+V_{12})V_2^{-1}
    \end{equation*}
where $Q \geq 0$ is the solution to the ARE:
    \begin{multline*} 
            (A-V_{12}V_2^{-1}C)Q+Q(A-V_{12}V_2^{-1}C)^T \\
            -QC^TV_2^{-1}Q+V_1-V_{12}V_2^{-1}V_{12}^T = 0.
    \end{multline*}
Note that, 
    \begin{equation*} 
        \begin{split}
            \mathbb{E} &
            \begin{bmatrix}
                B_u & B_w \\
                D_u & D_w
            \end{bmatrix}
            \begin{bmatrix}
                du \\
                dw
            \end{bmatrix}
            \begin{bmatrix}
                du \\
                dw
            \end{bmatrix}^T
            \begin{bmatrix}
                B_u & B_w \\
                D_u & D_w
            \end{bmatrix}^T \\
            & = 
            \begin{bmatrix}
                V_1 & V_{12} \\
                V_{12}^T & V_2
            \end{bmatrix} dt.
        \end{split}
    \end{equation*}
Next, we obtain a coherent LQG controller of the form~(\ref{controller}) by applying the appropriate method from \cite{ThienVuglarPetersen} or \cite{ MaaloufPetersen} based on the classical controller~(\ref{classiccontroller}) with $A_k$, $B_y$, and $C_k$ calculated above. To evaluate the cost~(\ref{cost}) explicitly, we consider the closed loop system:
\begin{equation}\label{closedloopsystem}
    d\zeta (t) = A_{cl}\zeta (t)dt+B_{cl}dw_{cl}(t); 
\end{equation}
where
\begin{equation*}
    \zeta =\begin{bmatrix} 
                x \\
                x_k
            \end{bmatrix}; \quad
    w_{cl} = \begin{bmatrix} 
                dw_1 \\
                dv_1 \\
                dv_2
            \end{bmatrix}
\end{equation*}
and 
\begin{equation}\label{clcost}
    J_{cl} = Tr(\bar{R}\bar{Q})
\end{equation} where $\bar{Q}$ is the unique symmetric positive definite solution of the Lyapunov equation
\begin{equation*}
    A\bar{Q}+\bar{Q}A^T+BB^T = 0; 
\end{equation*}
and
\begin{equation*}
    \bar{R} = \begin{bmatrix} 
                R_1 & 0 \\
                0 & C_k^TR_2C_k
            \end{bmatrix}.
\end{equation*}
That is, the cost function~(\ref{cost}) is evaluated using the expression~(\ref{clcost}). 

Our proposed algorithm can be summarized as follows:
\begin{enumerate}
    \item Beginning with matrices $A$, $B_{\hat{u}}$ and $C$ in~(\ref{plant}), we design a classical LQG controller~(\ref{classiccontroller}) using the standard approach~\cite[Theorem 5]{KwakSivan} and obtain $A_k$, $B_y$ and $C_k$.
    \item Implement~(\ref{classiccontroller}) as a coherent quantum controller using Theorem~\ref{Maaloufmethod} or~\cite[Section 4]{ThienVuglarPetersen}.
    \item Form the closed loop system~(\ref{closedloopsystem}) and evaluate the cost function~(\ref{clcost}).
\end{enumerate}
%%%%%%%%%%%%%%%%%%%%%%%%%%%%%%%%%%%%%%%%%%%%%%%%%%%%%%%%%%%%%%%%%%
\section{Example of an Equalization System}\label{example}
\begin{figure}[htp!] 
    \centering
    \includegraphics[width=8cm]{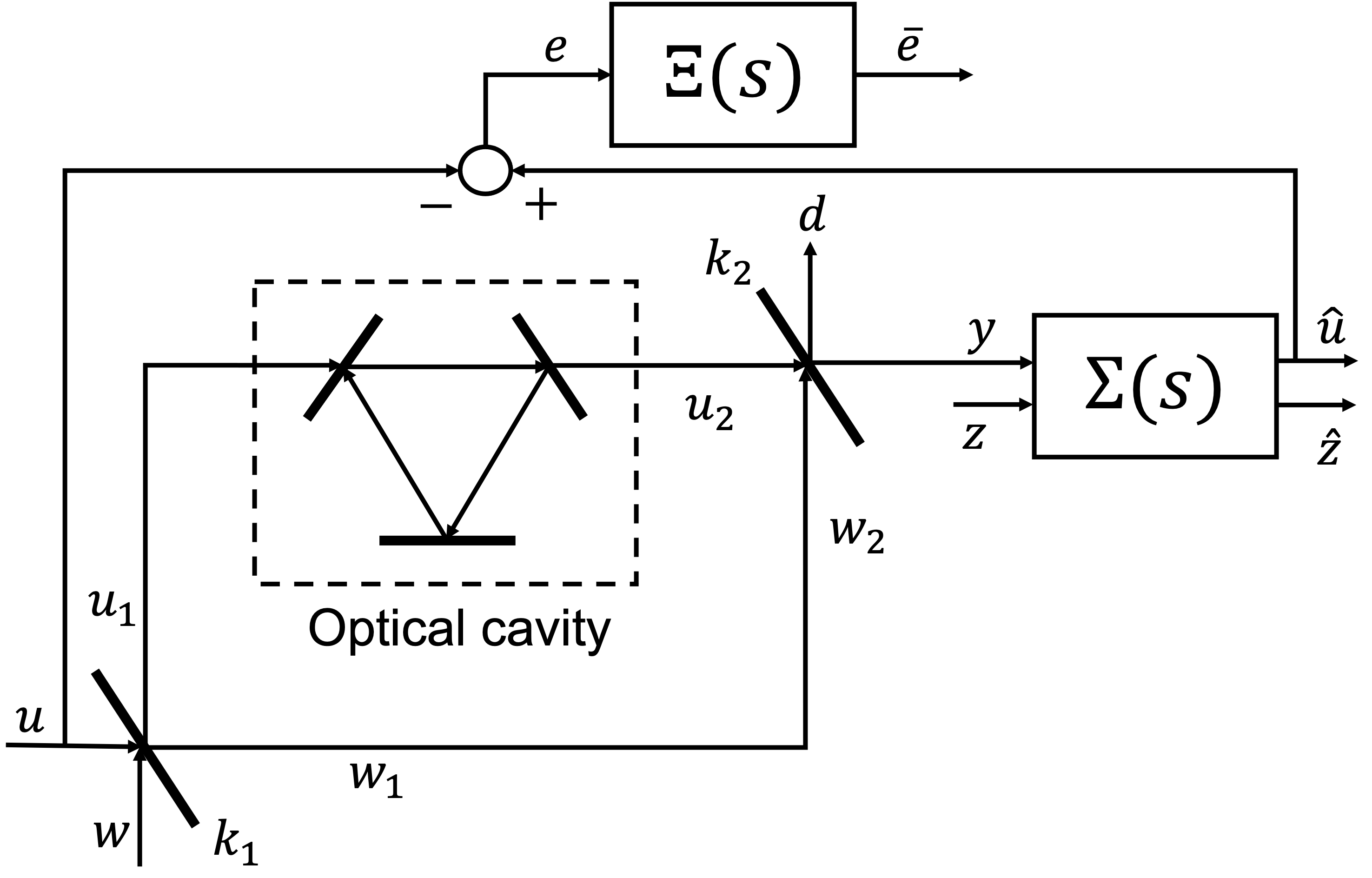}
    \caption{Equalization of an optical cavity system}
    \label{fig:system1}
\end{figure}
We now consider a modified example of an equalization system from~\cite{UgrinovskiiJames}, as shown in Figure~\ref{fig:system1}. The channel consists of an optical cavity and two optical beam splitters. The following are the constants used
\begin{equation*}
    \kappa = 5, \quad k = 0.4, \quad m = \sqrt{1-k^2}, \quad \Omega = 10~\text{and}~\tau = 0.1.
\end{equation*}
The constants are adapted from \cite[Section 6.2]{UgrinovskiiJames}.
We will consider both passive and active systems for the equalization filter~$\Sigma(s)$ in subsections~(\ref{expassive}) and~(\ref{exactive}), respectively. We will then comment on their relative performance. 

\begin{figure}[htp!] 
    \centering
    \includegraphics[width=8.5cm]{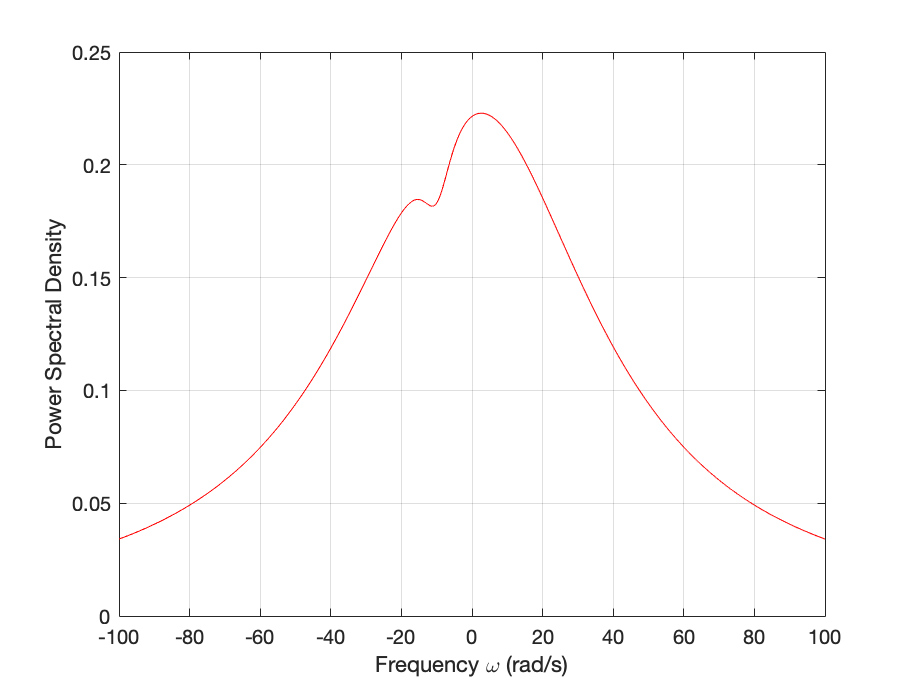}
    \caption{Closed loop Power Spectral Density for Passive Controller Design}
    \label{fig:passive}
\end{figure}

\subsection{Coherent LQG Control of a Passive Quantum System}\label{expassive}
Here our plant is of the form~(\ref{specialcase}) with
\begin{subequations}
    \begin{align}
        A & =\begin{bmatrix} \label{A}
                -k+i\Omega & 0 \\
                0 & -\frac{1}{\tau}
            \end{bmatrix}; \\
        B_{\hat{u}} & =\begin{bmatrix} \label{Bhat}
                0 \\
                \frac{1}{\tau}
                \end{bmatrix}; \\
        B_{w_1} & =\begin{bmatrix} \label{Bw1}
                -k\sqrt{2\kappa} & -m\sqrt{2\kappa} \\
                -\frac{1}{\tau} & 0
                \end{bmatrix}; \\
        C & =\begin{bmatrix} \label{C}
                k\sqrt{2\kappa} & 0
            \end{bmatrix}; \\
        D_{w_1} & =\begin{bmatrix} \label{Dw1}
                k^2-m^2 & 2km
                \end{bmatrix}  
            \end{align}
\end{subequations} 
and we choose $R_1$, $R_2$, $\mu$ of~(\ref{cost}) to be 
\begin{equation*}
    R_1 =\begin{bmatrix} 
                0 & 0 \\
                0 & 1
            \end{bmatrix}; \quad
    R_2 = 1; \quad
    \mu = 0.1.
\end{equation*}
The evaluated cost function~(\ref{clcost}) is $18.05$ and this is reflected in Figure~\ref{fig:passive} which gives the closed loop power spectral density of the quantity $\begin{bmatrix}
    R_1^\frac{1}{2} \bar{e}\\
    R_2^\frac{1}{2} \hat{u}
\end{bmatrix}$.

\begin{figure}[htp!] 
    \centering
    \includegraphics[width=8.5cm]{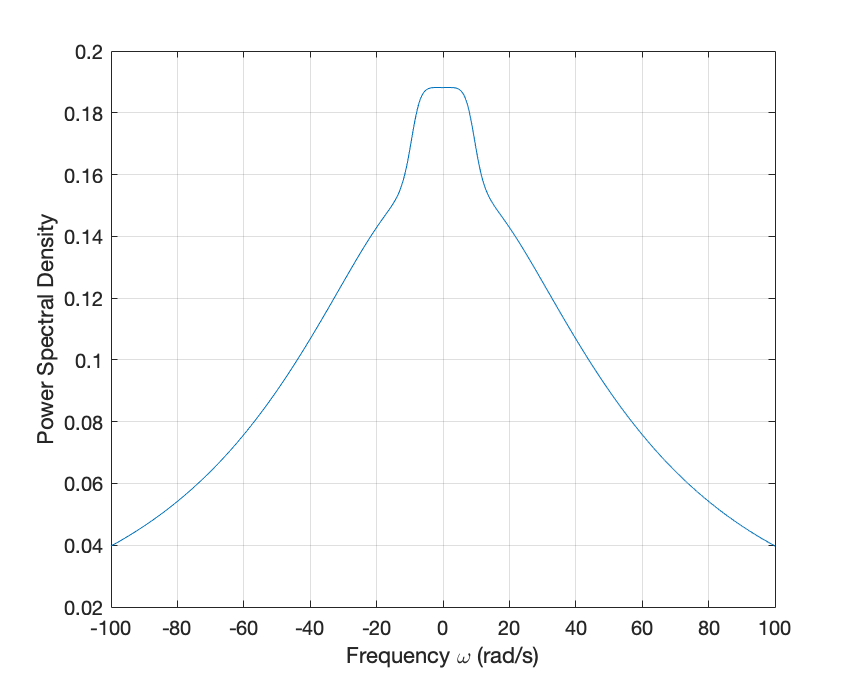}
    \caption{Closed loop Power Spectral Density for Active Controller Design}
    \label{fig:active}
\end{figure}

\subsection{Coherent LQG Control of an Active Quantum System} \label{exactive}
For the active case, we use a similar plant as in equations~(\ref{A}-\ref{Dw1}) obtained by applying the conversion matrix~\cite[Equation 22]{Petersen1}
\begin{equation*} 
    \Phi = \begin{bmatrix}
        I & I \\
        -iI & iI
    \end{bmatrix}
\end{equation*}
with appropriate dimensions to obtain matrices $A_g$, $B_{\hat{u}_g}$ and $C_g$ as follows:
    \begin{equation*}
        \begin{split}
            A_g & = \Phi A \Phi^{-1}; \\
            B_{\hat{u}_g} & = \Phi B_{\hat{u}} \Phi^{-1}; \\
            B_{{w_1}_g} & =\Phi B_{w_1}\Phi^{-1}; \\
            C_g & =\Phi C\Phi^{-1}; \\
            D_{{w_1}_g} & =\Phi D_{w_1}\Phi^{-1}
        \end{split}
    \end{equation*} 
and $R_1$, $R_2$, $\mu$ expands accordingly
    \begin{equation*}
        R_1 =\begin{bmatrix} 
                0_{2 \times 2} & 0_{2 \times 2} \\
                0_{2 \times 2} & I_{2 \times 2}
            \end{bmatrix}; \quad
        R_2 = I_{2 \times 2}; \quad
        \mu = 0.1.
    \end{equation*}
Now, the evaluated cost function~(\ref{clcost}) is $16.17$ and this is reflected in Figure~\ref{fig:active} which gives the closed loop power spectral density of the quantity $\begin{bmatrix}
    R_1^\frac{1}{2} \bar{e}\\
    R_2^\frac{1}{2} \hat{u}
\end{bmatrix}$.
% \begin{equation*}
%     \begin{split}
%         A & =\begin{bmatrix} 
%                 -k+i\omega & 10 & 0 & 0 \\
%                 -10 & -k+i\omega & 0 & 0 \\
%                 0 & 0 & -10 & 0 \\
%                 0 & 0 & 0 & -10 
%             \end{bmatrix}; \\
%         B_{\hat{u}} & =\begin{bmatrix} 
%                 0 & 0 \\
%                 0 & 0 \\
%                 10 & 0 \\
%                 0 & 10
%                 \end{bmatrix}; \\
%         B_{w_1} & =\begin{bmatrix} 
%                 -1.265 & -2.898 & 0 & 0 \\
%                 0 & 0 & -1.2649 & -2.898 \\
%                 -10 & 0 & 0 & 0 \\
%                 0 & -10 & 0 & 0 
%                 \end{bmatrix}; \\
%         C & =\begin{bmatrix} 
%                 1.265 & 0 & 1 & 0\\
%                 0 & 1.265 & 0 & 1
%             \end{bmatrix}; \\
%         D_{w_1} & =\begin{bmatrix} 
%                 -0.680 & 0.733 & 0 & 0 \\
%                 0 & 0 & -0.680 & 0.733
%                 \end{bmatrix}.  
%             \end{split}
% \end{equation*} 
\begin{figure}[htp!] 
    \centering
    \includegraphics[width=8.5cm]{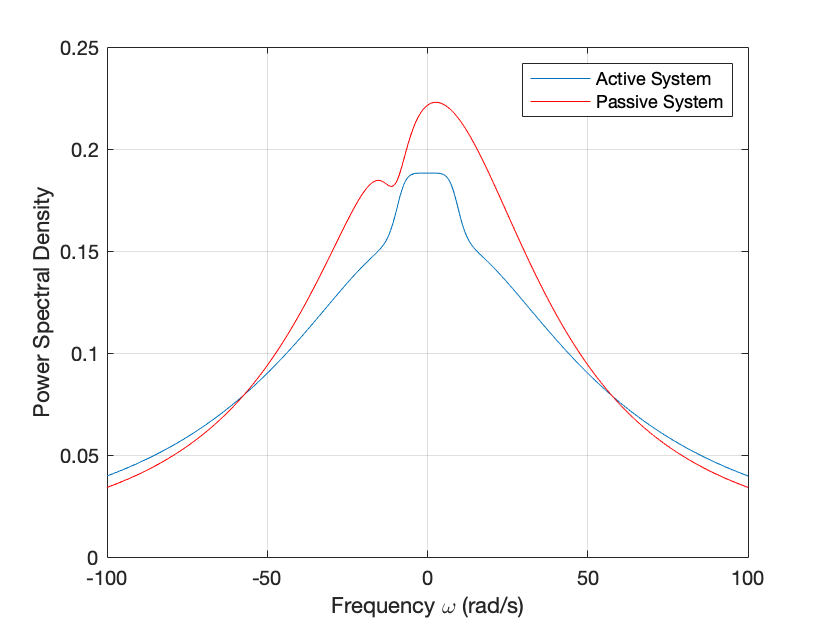}
    \caption{Closed loop Power Spectral Density for both Active and Passive Controller Designs}
    \label{fig:psd}
\end{figure}
\subsection{Comparison of controller system performance}
The performance of the passive and active system's cost function~(\ref{clcost}) is illustrated in Figure~\ref{fig:psd} in terms of the power spectral density graph. For this example, the active system gave only marginal improvement in comparison to the passive system. Note that this is consistent with the idea that the active system will perform at least as well as the passive system since the passive system is a special case of the active system.
%%%%%%%%%%%%%%%%%%%%%%%%%%%%%%%%%%%%%%%%%%%%%%%%%%%%%%%%%%%%%%%%%%%%%%%%%%%%%%%%
\section{CONCLUSION AND FUTURE WORK}\label{conclusion}

\subsection{Conclusion}
The general idea of quantum equalization is to design a feedback controller that is a physically realizable quantum system and can compensate for the error in a quantum communication channel.
In this work, we have proposed a method to find a physically realizable coherent LQG quantum controller that minimizes a cost function related to the system equalization error. Examples are shown for both passive and active linear quantum equalizers. 

\subsection{Future Work}
In the example section, the active coherent filter performed only marginally better than the passive coherent filter. In a typical experimental setup, such marginal performance gain may or may not justify the additional complexity of an active coherent filter. Future work will explore experimental validation of this work.
%%%%%%%%%%%%%%%%%%%%%%%%%%%%%%%%%%%%%%%%%%%%%%%%%%%%%%%%%%%%%%%%%%%%%%%%%%%%%%%%


\begin{thebibliography}{99}

\bibitem{Gregory}
E. B. Gregory, "Introduction," in Channel Equalization for Wireless Communications: From Concepts to Detailed Mathematics: IEEE, 2011, pp. 1-29.

\bibitem{UgrinovskiiJames1}
V. Ugrinovskii and M. R. James, "Wiener Filtering for Passive Linear Quantum Systems," 2019: American Automatic Control Council, pp. 5372-5377. 

\bibitem{UgrinovskiiJames} 
V. Ugrinovskii and M. James, "Coherent Equalization of Linear Quantum Systems," arXiv preprint arXiv:2211.06003, 2022.

\bibitem{GardinerZoller}
C. Gardiner and P. Zoller, {\it Quantum Noise}, Springer-Verlag Berlin Heidelberg; 2004.

\bibitem{BachorRalph}
Hans‐A. Bachor and T. C. Ralph, {\it A Guide to Experiments in Quantum Optics}, Wiley-VCH; 2004.

\bibitem{WallsMilburn}
D.F. Walls and G. J. Milburn, {\it Quantum Optics}, Springer-Verlag Berlin Heidelberg; 2008.

\bibitem{JamesNurdinPetersen}
M. R. James, H. I. Nurdin, and I. R. Petersen, "$H^{\infty}$ Control of Linear Quantum Stochastic Systems," IEEE Transactions on Automatic Control, vol. 53, no. 8, pp. 1787-1803, 2008.

\bibitem{VuglarPetersen}
S. L. Vuglar and I. R. Petersen, "Quantum Noises, Physical Realizability and Coherent Quantum Feedback Control," IEEE Transactions on Automatic Control, vol. 62, no. 2, pp. 998-1003, 2017.

\bibitem{MaaloufPetersen}
A. I. Maalouf and I. R. Petersen, "Coherent $H^{\infty }$ Control for a Class of Annihilation Operator Linear Quantum Systems," IEEE Transactions on Automatic Control, vol. 56, no. 2, pp. 309-319, 2011.

\bibitem{NurdinJamesPetersen} 
H. I. Nurdin, M. R. James, and I. R. Petersen, "Coherent quantum LQG control," Automatica, vol. 45, no. 8, pp. 1837-1846, 2009.

\bibitem{XiangPetersenDong}
C. Xiang, I. R. Petersen, and D. Dong, "Guaranteed cost coherent control for quantum systems with non-quadratic perturbations in the system Hamiltonian," Automatica, vol. 139, p. 110201.

\bibitem{DongPetersen}
D. Dong and I. R. Petersen, "Quantum estimation, control and learning: Opportunities and challenges," Annual Reviews in Control, vol. 54, pp. 243-251.

\bibitem{ThienVuglarPetersen}
R. T. Y. Thien, S. L. Vuglar, and I. R. Petersen, "Optimal Quantum Realization of a Classical Linear System," IFAC-PapersOnLine, vol. 53, no. 2, pp. 257-262, 2020/01/01/ 2020.

\bibitem{HudsonParthasarathy}
R. L. Hudson and K.R. Parthasarathy, Quantum \text{I}to's formula and stochastic evolutions, {\it Communications in Mathematical Physics}, vol. 93, 1984, pp 301-323.

\bibitem{Parthasarathy}
K.R. Parthasarathy, {\it An Introduction to Quantum Stochastic Calculus}, Birkhäuser Basel; 1992.

\bibitem{Belavkin}
V. P. Belavkin, Quantum continual measurements and a posteriori collapse on \text{CCR}, {\it Communications in Mathematical Physics}, vol. 146, 1992, pp 611-631.

\bibitem{KwakSivan}
H. Kwakernaak and R. Sivan, {\it Linear Optimal Control Systems}, Wiley Interscience; 1972.

\bibitem{BoutenHandelJames}
L. Bouten, R. Van Handel, and M. R. James, "An Introduction to Quantum Filtering," SIAM Journal on Control and Optimization, vol. 46, no. 6, pp. 2199-2241, 2007.

\bibitem{Petersen1}
I. R. Petersen, "Quantum Linear Systems Theory," The Open Automation and Control Systems Journal, vol. 8, pp. 67-93, 2016.

\bibitem{MaaloufPetersen1}
A. I. Maalouf and I. R. Petersen, "Bounded Real Properties for a Class of Annihilation-Operator Linear Quantum Systems," IEEE Transactions on Automatic Control, vol. 56, no. 4, pp. 786-801, 2011

\end{thebibliography}
\end{document}